\documentclass[12pt,a4paper]{article}
\usepackage{esint}
\usepackage{graphicx}
\usepackage{authblk}
\usepackage[a4paper,left=2.7cm,right=2.7cm,top=2.7cm,bottom=2.7cm]{geometry}
\begin{document}
	\providecommand{\keywords}[1]
	{
		\small	
		\textbf{\textit{Keywords---}} #1
	}
\title{Preisach Based Storage Devices and Global Optimizers}
%
\author[1]{Isaak D. Mayergoyz} 
\affil[1]{Department of Electrical and Computer Engineering, University of Maryland, College Park, MD 20742, U.S.A.}
\author[2]{Can E. Korman}
\affil[2]{Department of Electrical and Computer Engineering, George Washington University, Washington, DC 20052, U.S.A.}
%
\date{}
\maketitle
\begin{abstract}
The Preisach model of hysteresis admits simple device realizations. It is suggested in the paper that these realizations can be utilized as unique data storage devices as well as analog global optimizers.
\end{abstract}
%
%
%

%

%

\section{Introduction}
The Preisach model has long been used as a mathematical model of hysteresis phenomena of various physical nature. The purpose of this paper is to demonstrate that the Preisach model may find other very interesting applications. Namely, if the Preisach model is realized as a device with interconnected rectangular loop elements, then such realizations can be utilized as novel data storage devices, as well as analog global optimizers.

The paper consists of three parts. In the first part, the basic selective facts related to the Preisach model are briefly summarized and its device realizations are described. In the second part, the use of such realizations as novel data storage devices is explained. Finally, the third part deals with the discussion of utilization of Preisach model-type devices as unique analog type global optimizers.

\section{The Basic Facts Related to the Preisach Model}
\begin{figure}[t]
\centering \includegraphics[width=0.4\textwidth]{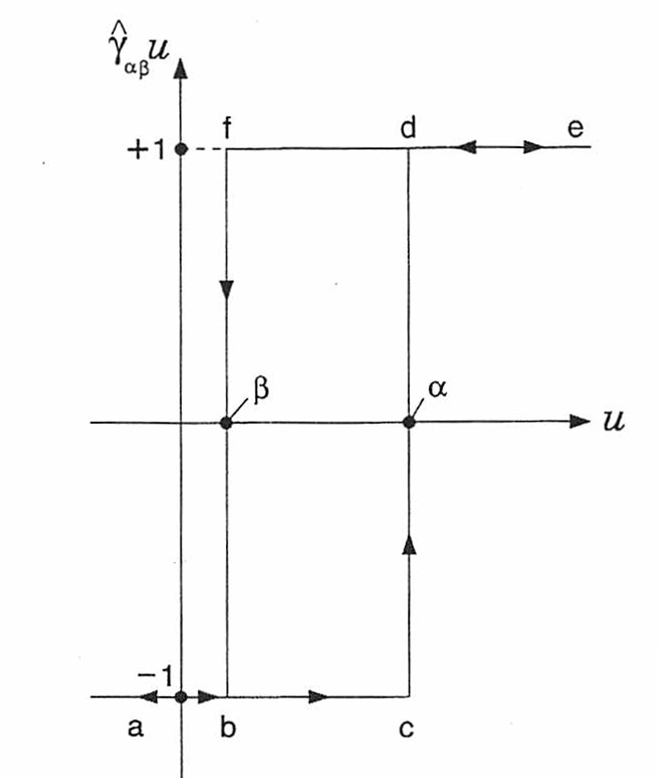}
\caption{Rectangular hysteresis loop operator.}
\label{recloop}
\end{figure}

The Preisach model is mathematically defined by the following formula \cite{IDM2003}-\cite{brokate}:
\begin{equation}
f(t) =  \iint\limits_{\alpha\geq\beta}
\mu (\alpha, \beta)
\hat{\gamma}_{\alpha, \beta} u(t) d\alpha d\beta
.
\end{equation}
Here: $u(t)$ and $f(t)$ can be viewed as input and output of the Preisach model, respectively; $\hat{\gamma}_{\alpha, \beta}$ are rectangular hysteresis loop operators (see Figure \ref{recloop}) with $\alpha$ and $\beta$ being the “up” and “down” switching thresholds, respectively, while $\mu (\alpha, \beta)$ is some weight function chosen by fitting to some experimental data during the model identification process.

It is apparent from the given definition that a complicated hysteresis operator is represented as a superposition of the simplest rectangular loop hysteresis operators. In this sense, formula (1) is mathematically similar to the spectral decomposition of self-adjoint operators in terms of very simple projection operators \cite{vonneumann}-\cite{friedman}. There is also an interesting parallel between the Preisach model and wavelet transforms \cite{daub}. Indeed, all rectangular loop operators $\hat{\gamma}_{\alpha, \beta}$ can be obtained by translation and dilation from the rectangular loop operator $\hat{\gamma}_{\pm 1}$, which can be regarded as the “mother loop operator.” Thus, the Preisach model can be viewed as a “wavelet operator transform.”

\begin{figure}[t]
\centering \includegraphics[width=0.5\textwidth]{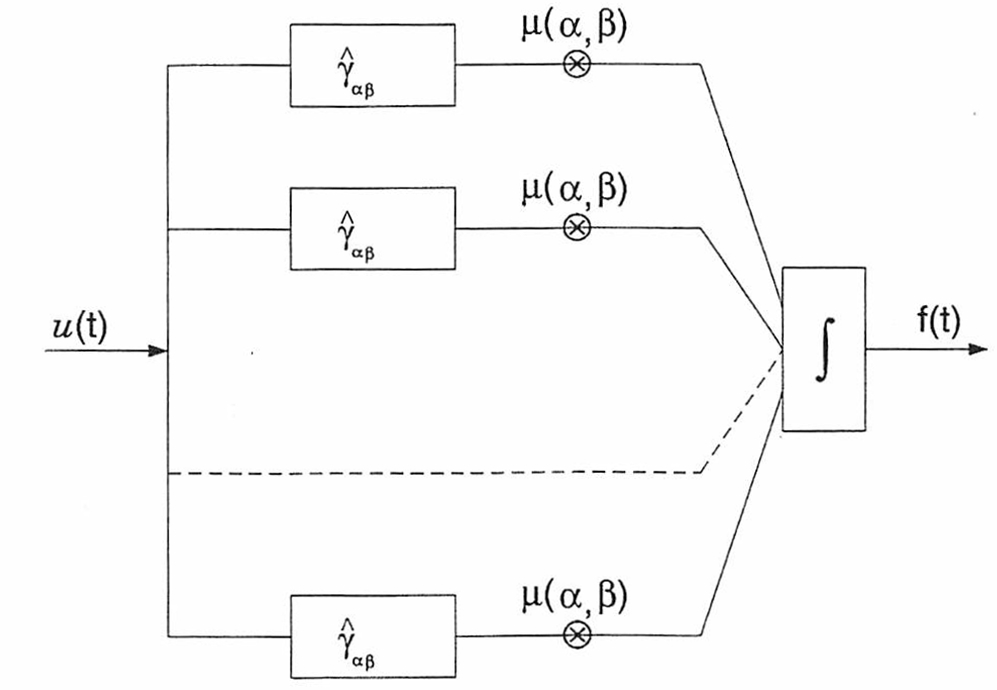}
\caption{Discrete approximation of the Preisach block-diagram.}
\label{preisach-block}
\end{figure}

It is apparent that the model (1) can be interpreted as a continuous analog of a system of parallelly connected rectangular loop elements. This interpretation is illustrated in Figure \ref{preisach-block}. According to this figure, the same input $u(t)$ is applied to all rectangular loop elements, their individual outputs are multiplied by  $\mu (\alpha, \beta)$, and then integrated over appropriate values of $\alpha$ and $\beta$. Discrete approximations of the block-diagram shown in Figure \ref{preisach-block} can be used for the construction of device realizations of the Preisach model.

\begin{figure}[t]
\centering \includegraphics[width=0.7\textwidth]{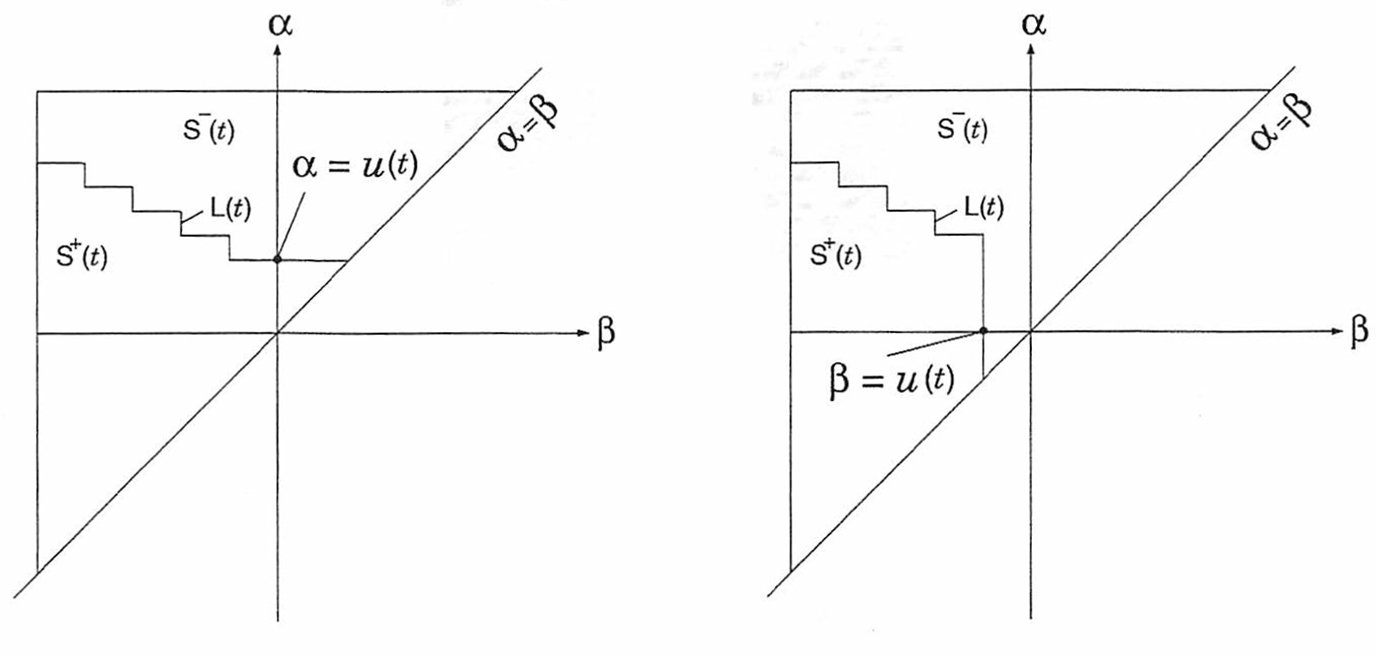}
\caption{Staircase interface on $(\alpha ,\beta )$ plane.}
\label{staircase}
\end{figure}

The mathematical study of the Preisach model is facilitated by its geometric interpretation. This interpretation is based on the one-to-one correspondence between rectangular loop operators $\hat{\gamma}_{\alpha, \beta}$ and points $(\alpha ,\beta )$ of the half-plane $\alpha \geq \beta$. It is also clear that at any instant of time this half-plane is subdivided into two sets: $S^+ (t)$ consisting of the points $\hat{\gamma}_{\alpha, \beta}$ for which $\hat{\gamma}_{\alpha, \beta} u(t) = 1$, and $S^- (t)$ consisting of the points $(\alpha ,\beta )$ for which $\hat{\gamma}_{\alpha, \beta} u(t) = -1$ (see Figure \ref{staircase}). It can be easily shown that if the starting configuration is a staircase interface or if the input is started from positive or negative saturation, the interface $L(t)$ between $S^+ (t)$ and $S^- (t)$ is a staircase-type line, where vertices have $\alpha$ and $\beta$ coordinates coinciding respectively with local maxima and minima reached at previous instants of time. The final link of $L(t)$ is attached to the line $\alpha = \beta$. This link is a horizontal one and it moves upwards, when the input is monotonically increased.  On the other hand, the final link is a vertical one and it moves from right to left, when the input is monotonically decreased. Thus, according to the formula (1), at any instant of time we have
\begin{equation}
f(t) =  \iint\limits_{S^+ (t)}
\mu (\alpha, \beta) d\alpha d\beta
-
\iint\limits_{S^- (t)}
\mu (\alpha, \beta) d\alpha d\beta
.
\end{equation}

\begin{figure}[t]
\centering \includegraphics[width=0.7\textwidth]{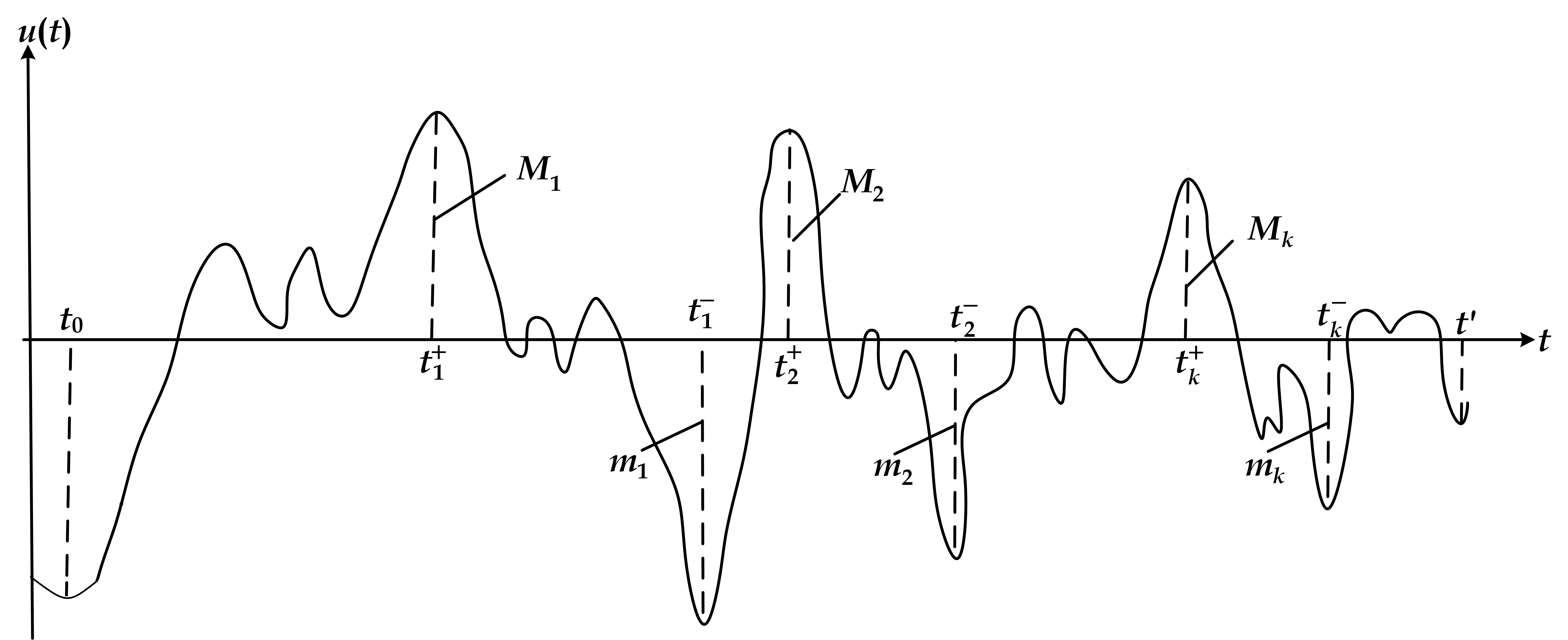}
\caption{Past input extrema selectively stored by the Preisach model.}
\label{extrema}
\end{figure}
It follows from formula (2) that instantaneous output value of $f(t)$ depends on a particular subdivision of the region of integration in (1) into positive $(S^+ (t))$ and negative $(S^- (t))$  sets. This subdivision is determined by  $L(t)$, whose shape is controlled by past extremum values of input $u(t)$. Consequently, the Preisach model is endowed with memory and $L(t)$ can be viewed as a trace (engram) of this memory. This memory is highly selective. Namely, the Preisach model detects, extracts and stores not all input values but only past input extrema. Furthermore, some of the stored input extrema can be erased by the subsequent input variations. This is illustrated by Figure \ref{extrema}, which shows that only specific sequences of input extrema are stored by the Preisach model and that these extrema are defined by the following formula:
\begin{equation}
M_1 = u(t_{1}^{+} ) = \max\limits_{t_{0} < t < t^{'}} u(t)
,
\end{equation}
\begin{equation}
m_1 = u(t_{1}^{-} ) = \min\limits_{t_{1}^{+} < t < t^{'}} u(t)
,
\end{equation}
\begin{equation}
M_2 = u(t_{2}^{+} ) = \max\limits_{t_{1}^{-} < t < t^{'}} u(t)
,
\end{equation}
\begin{equation}
m_2 = u(t_{2}^{-} ) = \min\limits_{t_{2}^{+} < t < t^{'}} u(t)
,
\end{equation}
\begin{equation}
\dots\dots\dots\dots
\nonumber
\end{equation}
\begin{equation}
M_k = u(t_{k}^{+} ) = \max\limits_{t_{k-1}^{-} < t < t^{'}} u(t)
,
\end{equation}
\begin{equation}
m_k = u(t_{k}^{-} ) = \min\limits_{t_{k}^{+} < t < t^{'}} u(t)
.
\end{equation}
It can be seen from the geometric interpretation of the Preisach model that all other (intermediate) input extrema are erased. It is natural to say that $M_k$ and $m_k$ $(k=1,  2, \dots )$ form a sequence of alternating dominant input maxima and minima. This is the essence of the erasure property of the Preisach model that can be stated as follows:

{\bf\it Erasure Property}: Only the sequences of alternating dominant input extrema are stored by the Preisach model. All other input extrema are erased.

	It is evident from the stated erasure property that $\alpha$ and $\beta$ coordinates of vertices of engram $L(t)$ are equal to $M_k$ and $m_k$ $(k=1, 2, \dots )$, respectively.

\section{Preisach Based Data Storage}
The existing conventional data storage systems are based on the availability of identical microelectronic devices used for the storage of binary numbers. However, the relentless process of miniaturization has resulted in the transition of digital electronic devices from microscale to nanoscale. At the nanoscale, electronic devices are very susceptible to random dopant fluctuations which occur due to the random nature of ion implantation and diffusion. There are also random fluctuations of oxide thickness in MOSFET devices. All these random fluctuations lead to fabrication deviations that are strongly pronounced at the nanoscale and which appreciably affect the characteristics of digital semiconductor devices, \cite{shin}-\cite{zhang}. This makes the fabrication of almost identical storage elements very expensive and even unattainable in the case of their further miniaturization. The nonidentical nature of digital devices is detrimental to the existing principles of computer storage. On the other hand, the nonidentical nature of storage elements is beneficial for the design of Preisach-type memory devices. Actually, it is at the very foundation of the design of such devices. For this reason, this type of devices may find some applications in future data storage systems.

One of the main difficulties in the Preisach based data storage is that the data subject to storage must be represented as sequences of alternating maxima and minima of decreasing magnitude. Only in this case, the Preisach storage without any erasure. i.e. without any information loss, can be accomplished. This requires special formatting of the data subject to storage. This formatting is discussed below.

Suppose that we have a set of numbers subject to storage. First, we subdivide this set of numbers into two subsets of positive and negative numbers, respectively. Namely, numbers
\begin{equation}\label{9}
 x_1 , x_2 , \dots , x_n
\end{equation}
are positive, while numbers
\begin{equation}\label{9}
 y_1 , y_2 , \dots , y_m
\end{equation}
are negative.

\begin{figure}[t]
\centering \includegraphics[width=0.6\textwidth]{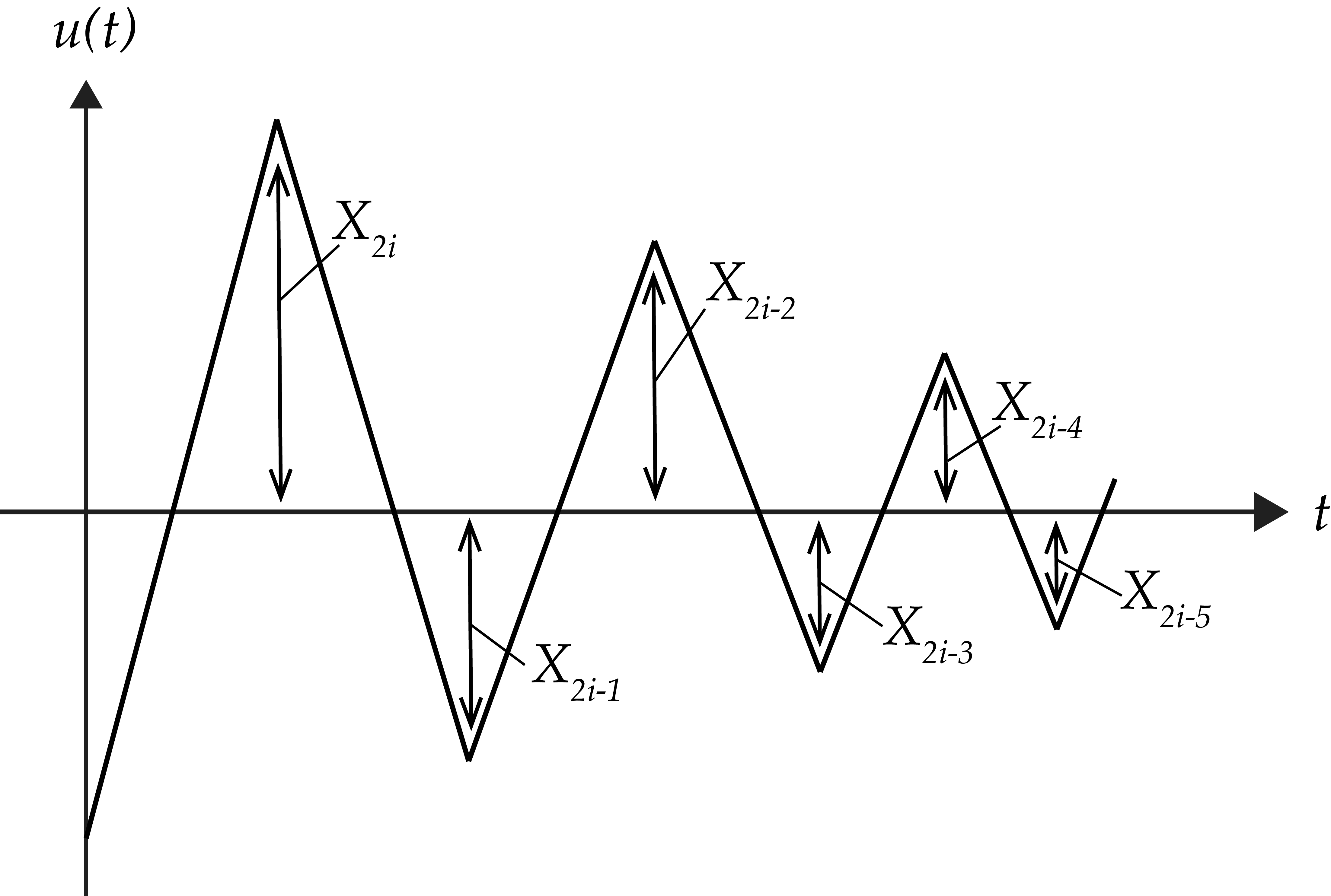}
\caption{Input sequence to the Preisach model.}
\label{zigzag}
\end{figure}

\begin{figure}[t]
\centering \includegraphics[width=0.6\textwidth]{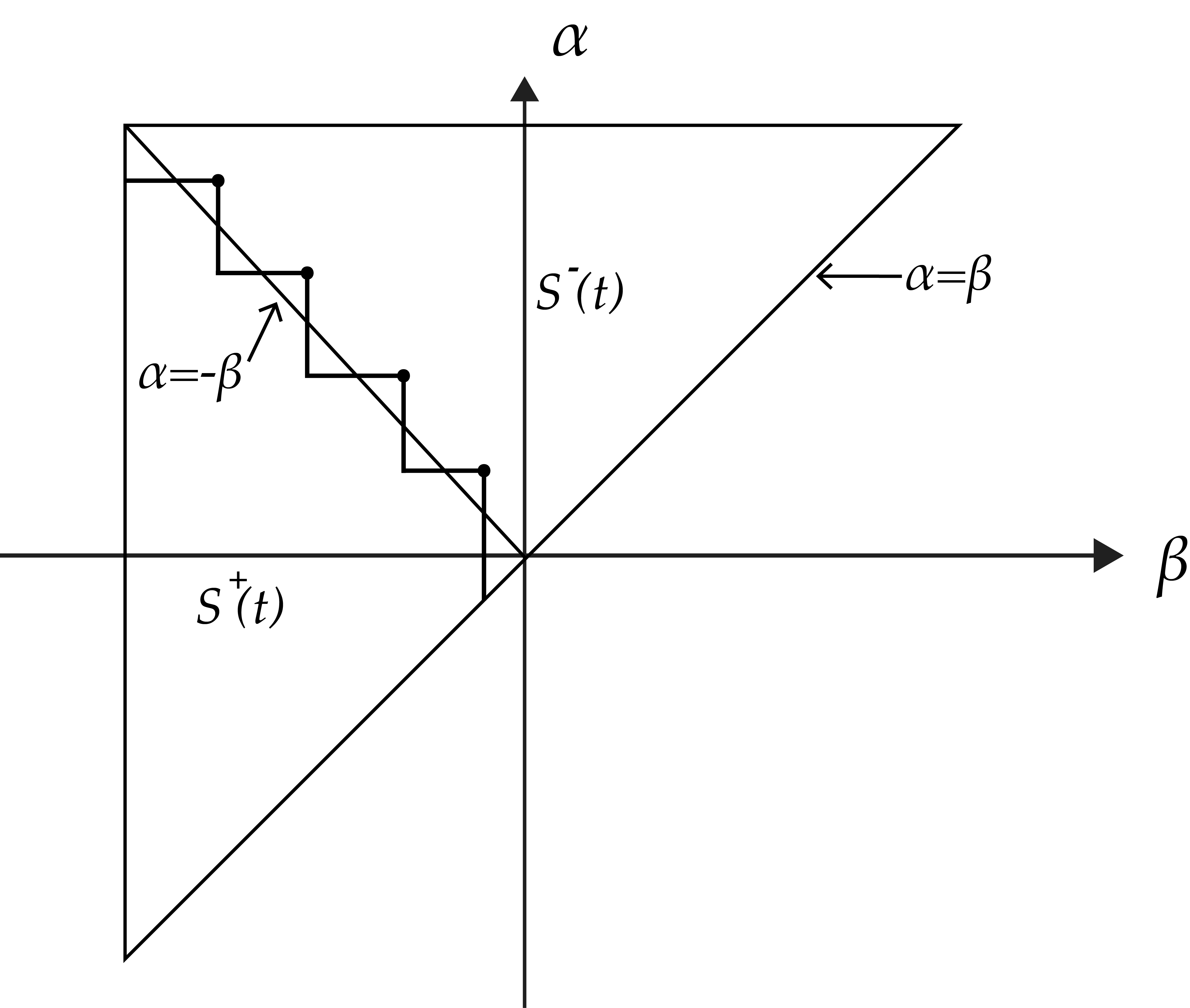}
\caption{Staircase interface of input sequence on the Preisach plane.}
\label{zigzagstair}
\end{figure}

Then, we introduce the following sequence of all positive numbers:
\begin{equation}
 x_1 , x_2 , \dots , x_n , x_{n+1} , x_{n+2} , \dots , x_N
 ,
\end{equation}
where
\begin{equation}
x_{n+1} = - y_1 , \ x_{n+2} = - y_2 , \dots , x_N = -y_m
\end{equation}
and
\begin{equation}
N=n+m
.
\end{equation}
Next, we design the following sequence of numbers $X_k$ in the case when $N=2i$:
\begin{equation}
X_{2i} = \sum_{k=1}^{2i} x_k , \
X_{2i-2} = \sum_{k=1}^{2i-2} x_k ,
\dots
,
X_{2} = \sum_{k=1}^{2} x_k  \
,
\end{equation}
\begin{equation}
X_{2i-1} = - \sum_{k=1}^{2i-1} x_k , \
X_{2i-3} = - \sum_{k=1}^{2i-3} x_k ,
\dots
,
X_{1} = - x_1  \
.
\end{equation}
From this definition, it follows that
\begin{equation}
X_{2i} >  X_{2i-2} > \dots > X_{2}
,
\end{equation}
and
\begin{equation}
X_{2i-1} <  X_{2i-3} < \dots < X_{1}
.
\end{equation}
In the case when $N=2i+1$, we define sequence $X_k$ as follows:
\begin{equation}
X_{2i+1} = \sum_{k=1}^{2i+1} x_k , \
X_{2i-1} = \sum_{k=1}^{2i-1} x_k ,
\dots
,
X_{1} =  x_1  \
,
\end{equation}
\begin{equation}
X_{2i} = - \sum_{k=1}^{2i} x_k , \
X_{2i-2} = - \sum_{k=1}^{2i-2} x_k ,
\dots
,
X_{2} = - \sum_{k=1}^{2} x_k  \
.
\end{equation}
Similarly, it follows that
\begin{equation}
X_{2i+1} >  X_{2i-1} > \dots > X_{1}
,
\end{equation}
while
\begin{equation}
X_{2i} <  X_{2i-2} < \dots < X_{2}
.
\end{equation}
The above defined sequences $X_k$ can be stored in the Preisach memory devices consisting of parallel connectivity of $\hat{\gamma}_{\alpha, \beta}$ without being affected by the erasure property of the Preisach memory. Indeed, by using an input shown in Figure \ref{zigzag}, we shall end up with the Preisach memory engram shown in Figure \ref{zigzagstair}.

The last figure corresponds to the case when $N=2i$. In this case, the last link of the engram connected to the line $\alpha = \beta$ is a vertical one. The $\alpha$ and $\beta$ coordinates of the vertices of the protruding angles of the engram are equal to $X_{2j}$ and $X_{2j+1}$, respectively.

In the case when $N=2i+1$, the engram will be similar to one shown in Figure \ref{zigzagstair}, except the last link of the engram will be a horizontal one and the $\alpha$ and $\beta$ coordinates of the vertices of the protruding angles of the engram are equal to $X_{2j+1}$ and $X_{2j}$, respectively.

As discussed below, the recorded data $X_k$ can be read back by monotonically increasing the input $u(t)$ and properly analyzing the output $f(t)$. As soon as numbers $X_k$ are read, the original numbers in formulas (9) and (10) can be recovered by using formulas (13), (14), and (15) in the case when $N=2i$ and formulas (12), (18), and (19) in the case when $N=2i+1$.

It can be seen from Figure \ref{zigzagstair} that the Preisach memory with only symmetric $(\alpha =- \beta )$ rectangular loops $\hat{\gamma}_{\alpha, \beta}$ can be used. Indeed, it follows from formulas (14)-(15) and (18)-(19) that the vertices of the protruding angles of the engram are above the line $\alpha =- \beta $, while the vertices of other (“caving in”) angles of the engram are below the line $\alpha =- \beta $. This fact follows from the inequality
\begin{equation}
X_{2j} > \vert X_{2j-1} \vert > X_{2j-2}
\end{equation}
in the case when $N=2i$, and the inequality
\begin{equation}
X_{2j+1} > \vert X_{2j} \vert > X_{2j-1}
\end{equation}
in the case when $N=2i+1$.

The output of the Preisach model in the case of symmetric rectangular loops can be written as follows
\begin{equation}
f(t) = C \int\limits_{\alpha}
\hat{\gamma}_{\alpha, -\alpha} u(t) d\alpha
,
\end{equation}
where it is assumed (for the sake of simplicity) that $\mu (\alpha, -\alpha) = C$.

\begin{figure}[t]
\centering \includegraphics[width=0.5\textwidth]{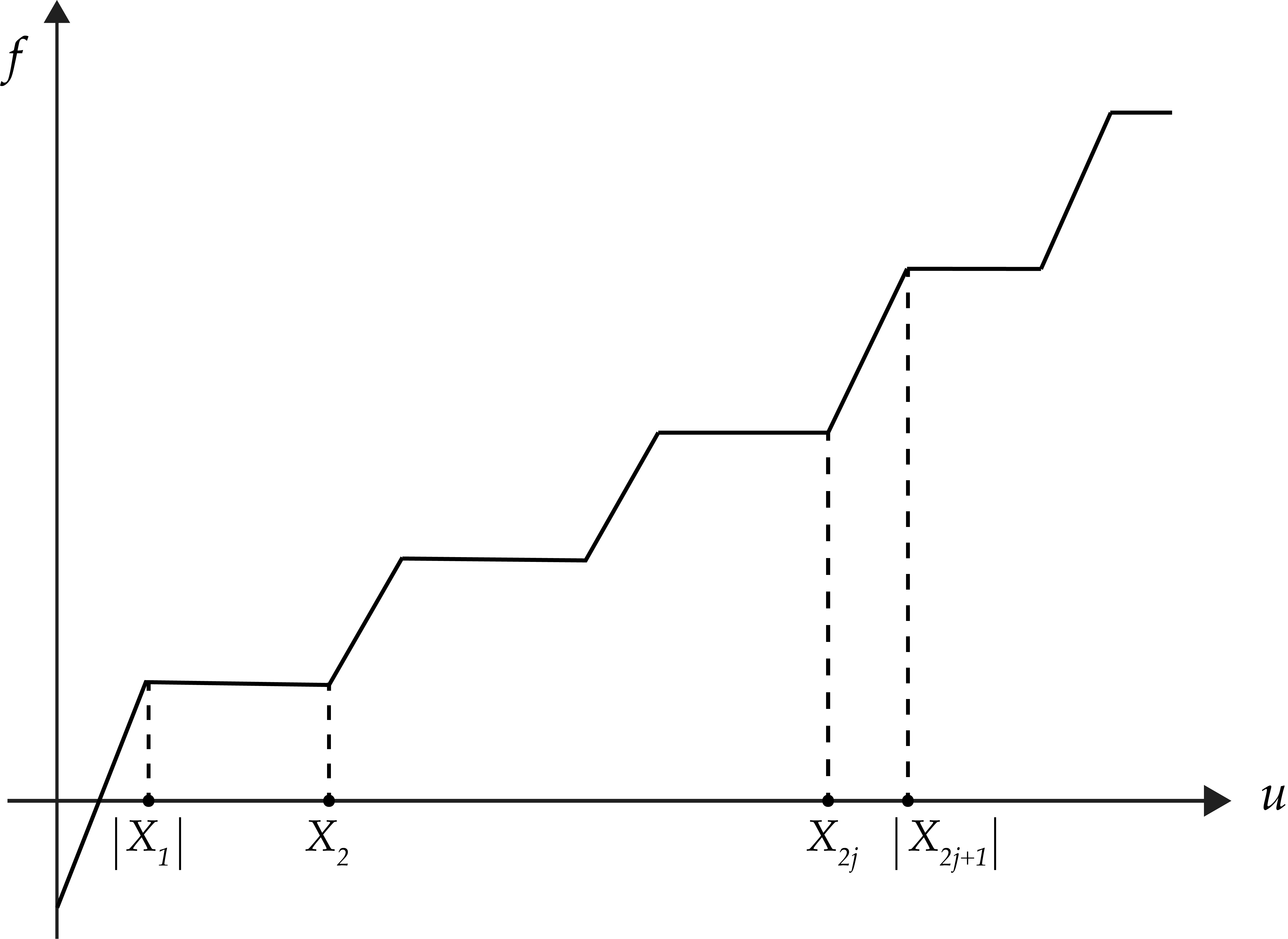}
\caption{Output $f$ as a function of input $u$.}
\label{risingoutput}
\end{figure}

Now, in the case of the reading back process when the input is monotonically increased, rectangular loops $\hat{\gamma}_{\alpha, -\alpha}$ corresponding to the points of the line $\alpha =- \beta $ inside $S^- (t)$ will be switched upwards. From Figure \ref{zigzagstair} it can be seen that the above points belong to the disjoint parts of the line $\alpha =- \beta $. When the switching of $\hat{\gamma}_{\alpha, -\alpha}$ along these parts occurs, then according to the formula (24) the output $f(t)$ is linearly increased as a function of input $u(t)$. Between these switchings, the output $f(t)$ remains constant. Furthermore, it follows from Figure \ref{zigzagstair} that the linear output increase starts at the input values $u(t_{2j} ) = X_{2j}$ and ends at the input values $u(t_{2j+1} ) =\vert  X_{2j+1} \vert$. This means that the output $f$ as the function of input $u$ can be represented by the graph shown in Figure \ref{risingoutput}.

This graph clearly reveals that the values of $X_k$ can be retrieved by using the described read back process. In the case when $\mu (\alpha, -\alpha)$  is not constant, the parts of the graph $f$ vs $u$ corresponding to the monotonic increase of output will not be linear but will be curved. However, these disjoint parts will be separated by horizontal (flat) parts whose initial and end points correspond to the input values equal to $\vert X_{2j-1} \vert$ and $X_{2j}$, respectively. Thus, the values of $X_k$  can still be recovered from the $f$ vs $u$ function.

Next, we shall discuss how the rectangular loop elements $\hat{\gamma}_{\alpha, -\alpha}$ can be physically realized. In microelectronics, this can be accomplished by using Schmitt trigger circuits with positive feedback. There are different realizations of such circuits by using BJT transistors or operational amplifiers. These realizations require also resistors. There are circuit realizations of the Schmitt trigger that can be accomplished without resistors. There are CMOS realizations, which use six MOSFETs \cite{filanovsky}-\cite{madikeri}. In magnetics, rectangular loop elements $\hat{\gamma}_{\alpha, -\alpha}$ can be realized by using nanoscale single-domain thin magnetic film particles with in-plane anisotropy. For such particles, the Stoner-Wolfarth theory can be used, which predicts rectangular magnetization loops when the applied magnetic field varies along the easy anisotropy axis \cite{IDM2003}. Finally, in optics, rectangular loop elements can be realized by using the physical phenomena of optical bistability. This phenomenon occurs in various semiconductors, and it can be also realized on a silicon chip \cite{almeida} despite the relatively weak nonlinear optical properties of silicon.

The discussed Preisach-type memory is of distributed nature, and, as a result, it is not addressable. For this reason, the most probable applications of this memory will be for massive data storage when all data stored in one Preisach memory unit is retrieved. Many identical Preisach memory units can be employed for massive data storage, and these units can be addressable.

\section{Preisach Based Global Optimizers}
Now, we turn to the discussion of the unique properties of the Preisach model-based global optimizers. Consider first the problem of univariate global optimization. The essence of this problem is in finding the global extremum (for instance, global minimum) of a given function $\varphi (t)$ of one variable:
\begin{equation}
\varphi ( \tilde{t} ) = \min\limits_{t \in [ 0,T ] } \varphi (t)
\end{equation}

Various numerical methods have been developed for the solution of this problem (for instance, see \cite{horst} -\cite{pardalos}). Usually, some smoothness of $\varphi (t)$ is assumed in these methods. One of the most commonly used assumptions is the differentiability of the objective function $\varphi$ with bounded derivatives (partial derivatives in the case of multivariate optimization) that leads to this function satisfying the Lipschitz condition, i.e.,
\begin{equation}
\vert
\varphi ( t_2 ) - \varphi ( t_1 )
\vert
\leq
L \vert t_2 - t_1 \vert
\end{equation}
where $L>0$ is the Lipschitz constant.

The multivariate global optimization problem is substantially more difficult. It requires the finding of the global minimum of the function of many variables
\begin{equation}
\varphi ( \vec{x^{*}} ) = \min\limits_{\vec{x} \in \mathcal{D} } \varphi (\vec{x})
\end{equation}
where $\vec{x} = ( x_1 , x_2 , \dots , x_n )$ and $\mathcal{D}$ is some region (for instance) a cube, in $n$-dimensional space.

It is natural to apply one-dimensional global optimization techniques to the solution of multidimensional global optimization problems. One of such approaches is the nested dimension reduction scheme, \cite{gergel}, \cite{grishagin}.

In this context, the Preisach model-based devices can be viewed as general global optimizers. Indeed, if the function $\varphi (t)$ subject to minimization is used as an input$\big( u(t) = \varphi (t) \big)$ to the Preisach based device, then this device will detect, extract, and store the alternating sequence of dominant extrema of $\varphi (t)$. Among these extrema are the global minima and maxima of $\varphi (t)$. The Preisach device does this due to its structure and no computations are needed if $\varphi (t)$ is given as an analog signal. The extracted and stored extrema of $\varphi (t)$ can be retrieved. This retrieval is especially simple in the case when the function $\mu (\alpha, \beta)$ is constant and the output of the Preisach memory is given by the formula:
\begin{equation}
f(t) =  C \iint\limits_{\alpha\geq\beta}
\hat{\gamma}_{\alpha, \beta} u(t) d\alpha d\beta
.
\end{equation}

\begin{figure}[t]
\centering \includegraphics[width=0.5\textwidth]{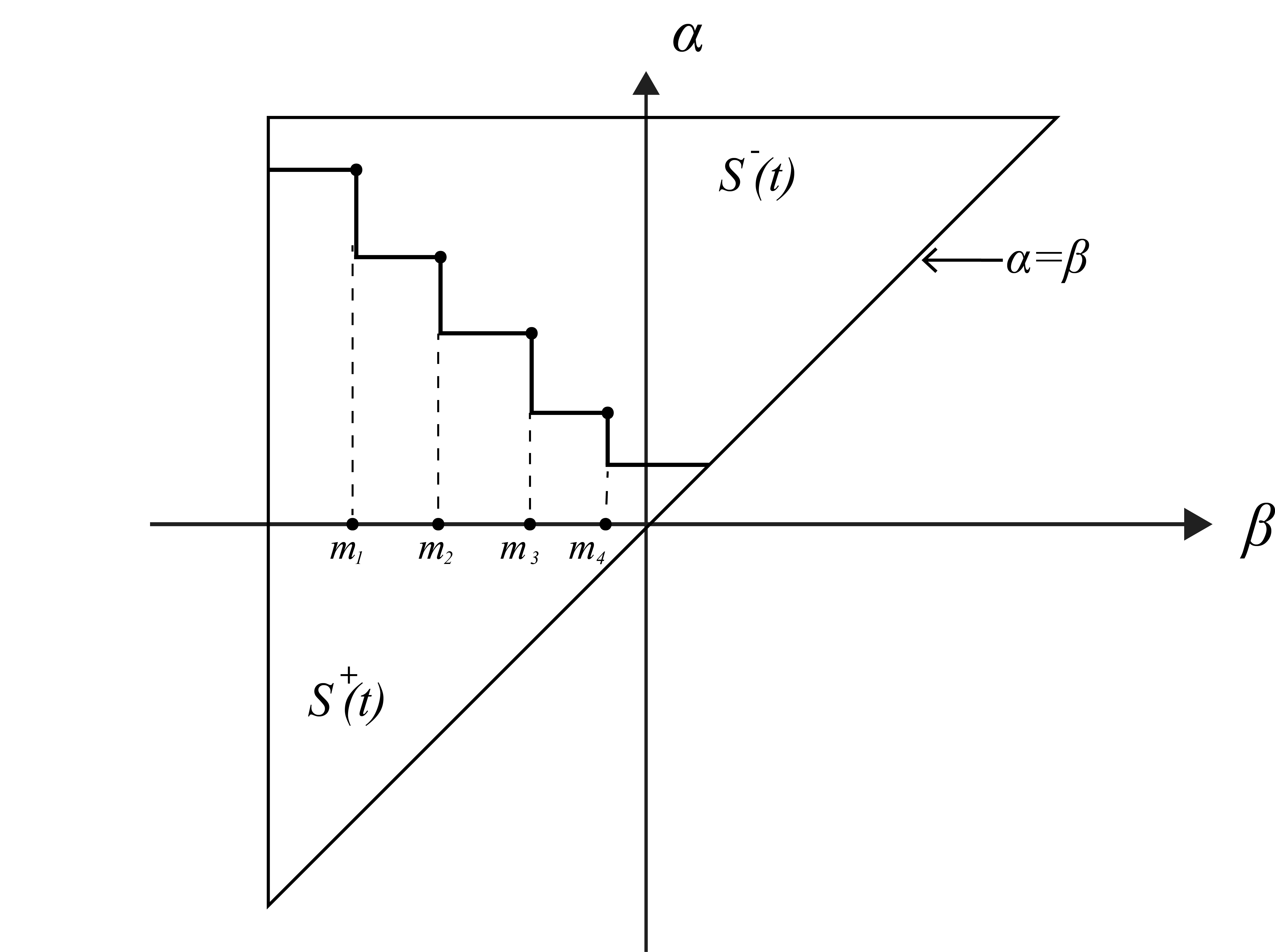}
\caption{Staircase interface on the Preisach plane.}
\label{staircase8}
\end{figure}

\begin{figure}[t]
\centering \includegraphics[width=0.5\textwidth]{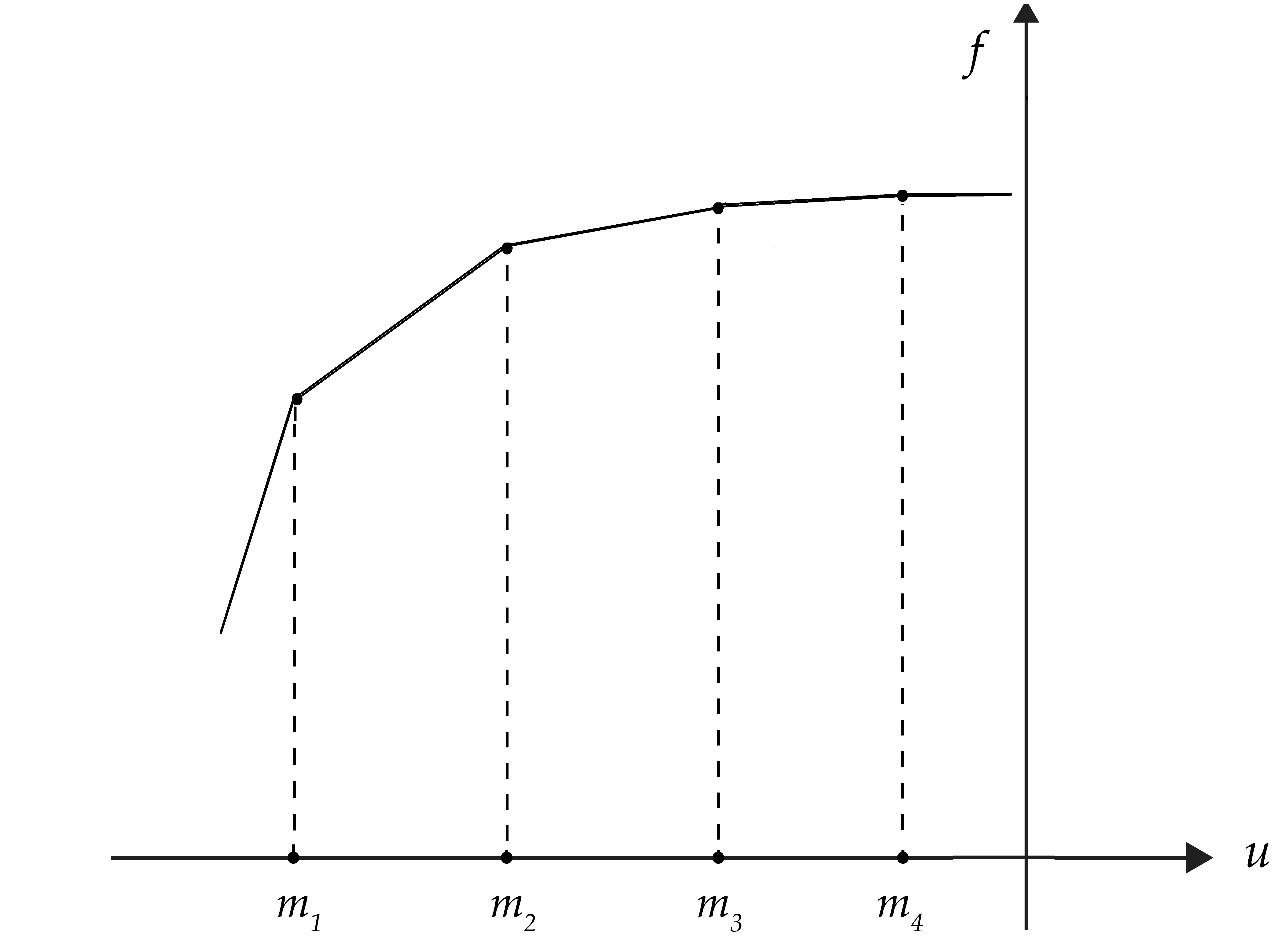}
\caption{Output $f$ as a function of input $u$.}
\label{fallingoutput}
\end{figure}

\noindent
Indeed, by using the diagram shown in Figure \ref{staircase8} and taking into account that the retrieval can be accomplished by using a monotonically decreasing input $u(t)$, it can be shown that the output $f(t)$ is a piece-wise linear function of $u(t)$ shown schematically in Figure \ref{fallingoutput}. The transition from one linear link of the function $f$ vs $u$ to another occurs at the input values $u=m_k$, where $m_k$ are extracted minima of $\varphi (t)$.

\begin{figure}[t]
\centering \includegraphics[width=0.5\textwidth]{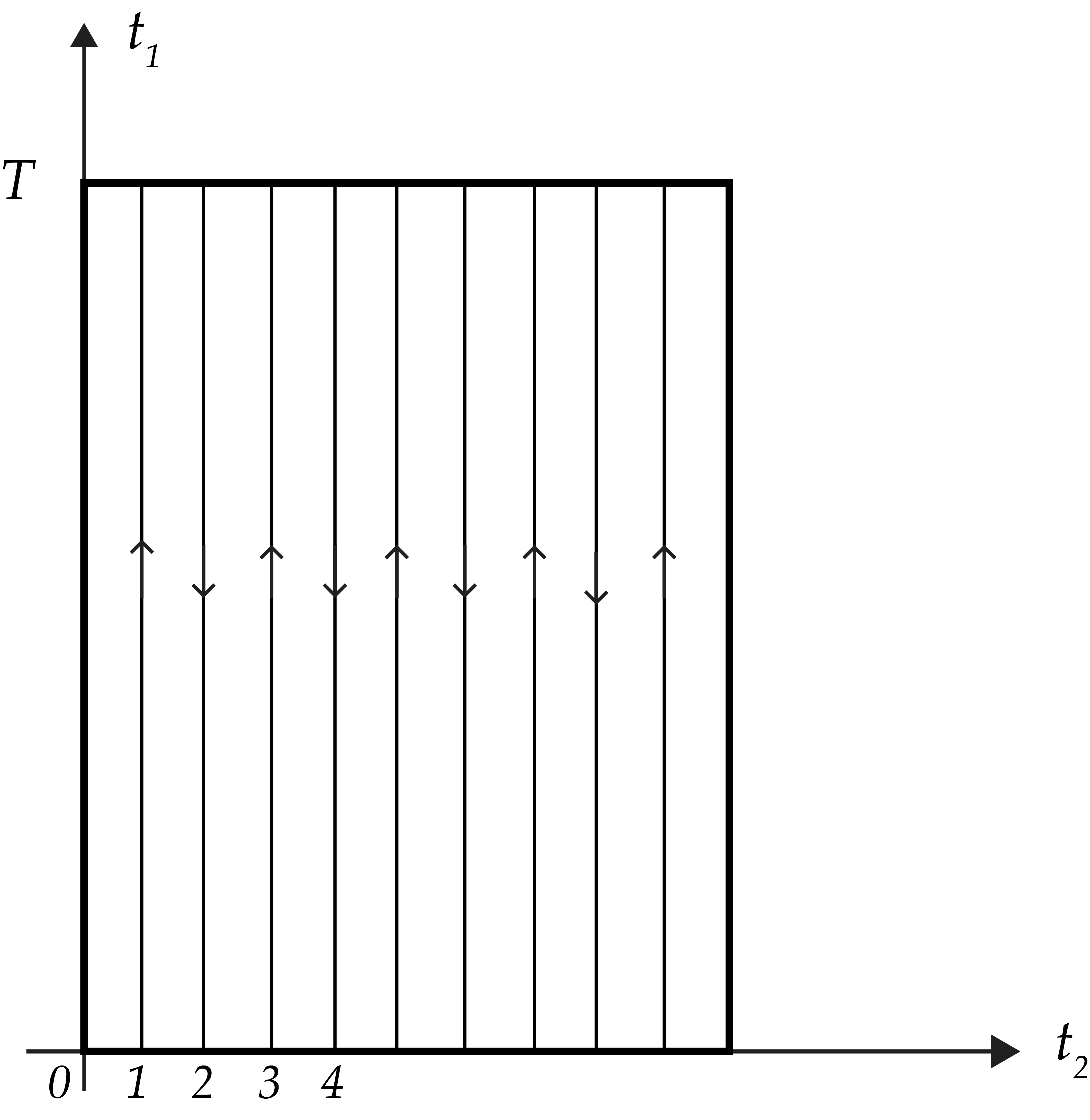}
\caption{Rectangular domain $\mathcal{D}$ illustrating a simple discretization.}
\label{verticles}
\end{figure}

The described approach can be applied to multivariate minimization, as well. Clearly, this is a broad problem. Therefore, in order to illustrate the approach, we shall limit our discussion here to a simple two-dimensional case where the domain $\mathcal{D}$ is a rectangle shown in Figure \ref{verticles}. Further, the simple nested dimension reduction scheme of finite discretization is employed in order to demonstrate that two-dimensional (or higher-dimensional) data can be used as scalar input to the Preisach operator.
\noindent
In this illustative case, the input $u(t)$ to the Preisach minimizer coincides with $\varphi (t)$  traced upwards along the vertical line $1$ during the interval $[ 0 , T ]$ then it coincides with $\varphi (t)$ traced downwards along the vertical line $2$ during the interval  $[ T , 2T ]$, and it is defined similarly for all other vertical lines. It is apparent that the Preisach minimizer will detect and store the global minima at all these vertical lines. This approach is very effective when the data along these vertical lines are available in analog form. This is the case when these data are obtained as a result of measurements. The described approach can be generalized to higher dimensions, and several Preisach minimizer units can be operated in parallel. It is apparent that some smoothness of $\varphi (\vec{x} )$ is tacitly assumed when minimization along discrete parallel lines is used. The same is true when finite number of $\hat{\gamma}$-elements (instead of infinite numbers) are used in realizations of the Preisach model (28). A more comprehensive analysis of dimension reduction, and the associated analysis for data smoothness and finite number of Preisach operators is left unattended in this paper.

\section{Conclusion}

By taking advantage of the inherent properties of the Preisach model and the fact that the model admits simple device realizations, it was shown in the paper that these realizations can be utilized as unique data storage devices, as well as analog global optimizers. In order to demonstrate these features, the Erasure Property of the Preisach model was employed, where only the sequences of alternating dominant input extrema are stored and where all other input extrema are erased.

For storage, it was shown that by specially formatting the input data, Preisach type storage devices can provide a robust mechanism to store and to extract information. The robustness is achieved due to the inherent distributed nature of the Preisach model where storage is due to the superposition of the outputs of large numbers of simple rectangular loops. It is important to point out here that the proposed Preisach based storage is distributed in nature with built-in data redundancy and differs substantially from the addressable storage architecture of conventional computing systems. In this context, this proposed storage and extraction mechanism presents an alternative robust architecture that may possess unique advantages for data back-up, as well as disadvantages due to its non-addressable nature. It is also important to point out here that the distributed and non-addressable nature of Preisach storage may also possess certain advantages in data security. Therefore, the benefits of its utilization as a primary vs. secondary storage mechanism in computing systems need to be explored further. However, the further exploration of these aspects is left as future work and would need additional study based on the requirements of diverse computing applications.

For global univariate optimization, it was pointed out that the Preisach device detects, extracts, and stores sequence of dominant extrema of the function optimized, and that among these extrema are the global minima and maxima of this function. It was emphasized that the Preisach device achieves this due to its inherent structure and that no additional computations are necessary if the input is an analog signal. Further, it was shown how the extracted and stored extrema can be retrieved by simple selection of the Preisach distribution function. In addition, some generalizations of this technique to multi-variate optimization were also suggested.

It is important to point out here that there are many methods available in the case of objective functions that have continuous partial derivatives. The presented approach has no such constraint, and it has the advantage that it allows optimization of Lipschitz functions that are not differentiable. It is clear that the topic of multi-variate optimization and alternative approaches to the generalization of univariate optimization requires additional careful analysis to explore this advantage further. However, the goal of this paper is to simply show the feasibility that global optimization can be achieved due to the inherent Erasure Property of the Preisach model for a broader range of functions compared to conventional techniques. With this in mind, the detailed analysis of univariate and many possible multi-variate optimization generalizations is left unattended in this paper and leaves these matters to future work.


\end{document}